\documentclass[a4paper]{jpconf}
\usepackage[utf8]{inputenc}
\usepackage{lineno}
\usepackage[czech,greek,german,british]{babel} 
\usepackage[T1]{fontenc} 
\usepackage{lmodern} 
\usepackage{graphicx}
\usepackage{xcolor}
\usepackage{amsthm}
\usepackage{amsmath} 
\usepackage{amssymb} 
\usepackage{tabularx} 
\usepackage{rotating} 
\usepackage{datetime} 
\usepackage{epstopdf} 
\usepackage{graphicx} 
\usepackage{booktabs} 
\usepackage[unicode]{hyperref} 
\usepackage{sidecap}
\usepackage{verbatimbox}
\sidecaptionvpos{figure}{t}

\newcommand{\num}[2]{#1\:{\textnormal{#2}}} 

\newcommand{\gr}[1]{{\foreignlanguage{greek}{#1}}} 

\newcommand{\urlss}[1]{\hbox{\scripTsize \url{#1}}} 

\newcommand{\pT}{\ensuremath{p_{\textup{T}}}} 
\newcommand{\pTj}{\ensuremath{{p^\textup{ch}_{\textup{T, jet}}}}}
\newcommand{\pTv}{\pT^\textup{V0}} 
\newcommand{\etav}{\eta^\textup{V0}} 
\newcommand{\kt}{\ensuremath{k_{\textup{t}}}} 
\newcommand{\akt}{anti-\kt} 
\newcommand{\mc}{\ensuremath{{\textup{MeV$/c$}}}} 
\newcommand{\gc}{\ensuremath{{\textup{GeV$/c$}}}} 
\newcommand{\tc}{\ensuremath{{\textup{TeV$/c$}}}} 

\newcommand{\snn}{\ensuremath{\sqrt{s_\textup{NN}}=\num{2.76}{TeV}}} 
\newcommand{\snnpPb}{\ensuremath{\sqrt{s_\textup{NN}}=\num{5.02}{TeV}}} 
\newcommand{\pbpb}{\textup{Pb--Pb}} 
\newcommand{\pp}{\textup{pp}} 
\newcommand{\ppb}{\textup{p--Pb}} 

\newcommand{\particle}[1]{\textup{#1}} 
\newcommand{\partgr}[1]{{\particle{\gr{#1}}}} 
\newcommand{\anti}[1]{{\ensuremath{\overline{#1}}}} 

\newcommand{\proton}{{\particle{p}}}

\newcommand{\pion}{\partgr{p}}

\newcommand{\kaon}{\particle{K}}

\newcommand{\kos}{\ensuremath{{\kaon^0_\textup{S}}}}
\newcommand{\lmb}{\partgr{L}}
\newcommand{\almb}{\anti{\partgr{L}}}

\newcommand{\decay}{\ensuremath{\rightarrow}} 

\begin{document}
\title{Production of strange particles in charged jets in $\textup{p--Pb}$ and $\pbpb$ collisions measured with ALICE at the LHC}

\author{Alice Zimmermann$^{1}$ (for the ALICE Collaboration)}
\address{1. Physikalisches Institut, Ruprecht-Karls-Universit{\"a}t Heidelberg, Im Neuenheimer Feld 226, 69120 Heidelberg, Germany}
\author{\href{mailto:alice.zimmermann@cern.ch}{\texttt{alice.zimmermann@cern.ch}}}

\begin{abstract}
Studies of jet production can provide information about the properties of the hot and dense strongly interacting matter created in ultra-relativistic heavy-ion collisions.
Specifically, measurement of strange particles in jets may clarify the role of fragmentation processes in the anomalous baryon to meson ratio at intermediate particle $p_\textup{T} $ that was observed in lead-lead (Pb--Pb) and, to a lesser extent, in proton-lead (p--Pb) collisions. 

In this contribution, measurements are presented of the $p_\textup{T}$ spectra of $\Lambda$($\overline{\Lambda}$) baryons and $\textup{K}^0_\textup{S}$ mesons produced in association with charged jets in Pb--Pb collisions at $\sqrt{s_\textup{NN}}=2.76\:\textup{TeV}$ and\\p--Pb collisions at $\sqrt{s_\textup{NN}}=5.02\:\textup{TeV}$. The analysis is based on data recorded by ALICE at the LHC, exploiting its excellent particle identification capabilities. The baryon/meson ratios of strange particles associated with jets are studied for different event activities in p-Pb collisions and are restricted to central events in Pb-Pb collisions. A comparison is shown to the ratios obtained for inclusive particle production and for particles stemming from the underlying event as well as to PYTHIA proton-proton (pp) simulations. 
\end{abstract}

\section{Introduction}

The first measurements of the baryon-meson ratio in heavy-ion collisions at the Relativistic Heavy Ion Collider (RHIC) showed an enhanced baryon/meson production at intermediate transverse momentum ($\pT$ = 3~$\gc$)~\cite{Agakishiev:2011ar,STAR2}
relative to $\pp$ collisions. Recent results at LHC energy corroborated this
observation~\cite{LFspectrapaper, BaryMesID}. There are several scenarios proposed to explain this, e.g. collective effects and string fragmentation in a hydrodynamically expanding environment~\cite{LaK0sHydro}. These collective phenomena~\cite{EllFlow2} are a characteristic feature of the Underlying Event (UE) in $\pbpb$ collisions, that are independent of the fragmentation process but could have a possible impact on the jet fragments. Interactions between partons, stemming from the fragmentation shower and the Quark-Gluon Plasma, could change the jet pattern, which was already observed as quenching of jets~\cite{JetQuench}. 
Some other models consider alternative hadronisation mechanisms, e.g. recombination at low $\pT$~\cite{CoalRecomb}, which is a soft process that could favour baryon over meson production. Particles with a momentum larger than $\pT$~=~4-6~$\gc$ would, on the other hand, be produced in hard processes via fragmentation, that does not lead to an enhanced production of baryons (compared to production in vacuum). 
 The study of identified particle yields and ratios like $\lmb$($\almb$) and $\kos$ in inclusive production and comparison to production in jets will allow to trace back possible differences in the hadronisation process and its $\pT$ dependence.
Since collective features were proposed to also occur in $\ppb$ collisions, one can compare among the different collision systems ($\pp$, $\ppb$ and $\pbpb$), to study the onset of this baryon enhancement, that has already been observed for inclusive production in $\ppb$ collisions~\cite{Abelev:2013haa}.
For the analyses presented in this contribution, the $\pT$ spectra of strange particles associated with jets are studied for events with different event multiplicities in $\ppb$ collisions and for the 10$\%$ most central events in $\pbpb$ collisions. 

\section{Experimental setup and analysis strategy}

The jet reconstruction is done with the $\akt$ jet finder using charged tracks with $\pT$~>~150~$\mc$ and requiring a leading particle in each jet to have at minimum $\pT$ = 5~$\gc$. In the following, jets fulfilling all these requirements are referred to as 'selected' jets. The jet energy is corrected on an event-by-event basis for the contribution of the average charged UE~\cite{JetEnSubtr}. 

Both $\ppb$ and $\pbpb$  analyses use the decay topology of the strange neutral particles (V0s) for their reconstruction. The decay channels  $\kos\decay\pion^++\pion^-$~\cite{pdg-la} and $\lmb\decay\proton+\pion^-$ ($\anti{\lmb}\decay\anti{\proton}+\pion^+$)~\cite{pdg-la} are used. By recombining the charged tracks of opposite curvature, measured in the Inner Tracking System (ITS) and the Time Projection Chamber (TPC), together with a particle mass hypothesis the invariant mass of the V0 candidates is calculated. A set of cuts applied on various observables of the topology of the decay already rejects a sizeable amount of combinatorial background.   
The particle signal is extracted after a background fit and by applying a bin-counting procedure on the invariant mass distributions in different intervals of $\pTv$ and $\pTj$.
The V0 candidate is associated to the jet cone if the distance, which is calculated in azimuthal angle ($\phi$) and in pseudo-rapidity ($\eta$), to the jet axis is smaller than the jet resolution parameter R:
\begin{equation}
\sqrt{(\phi_\text{V0}-\phi_\text{jet})^{2}+(\eta_\text{V0}-\eta_\text{jet})^{2}} < R 
\end{equation}
The V0 acceptance is chosen to $|\etav|$ < 0.7, the acceptance for the jets is $|\eta^\textup{jet}|$ < $|\etav| - R$  = 0.5(0.4) for R = 0.2(0.3).

\begin{SCfigure}
\centering
\caption{\label{fig:UncV0Jet} \addvbuffer[4ex 0ex] Uncorrected $\kos$ and $\lmb$ $\pT$ spectra for R = 0.3 jets in $\pbpb$ collisions at 2.76 TeV and in the 10$\%$ highest multiplicity class. The jets are reconstructed for two different jet $\pT$ intervals ($\pTj$ > 10 $\gc$, > 20 $\gc$) with the $\akt$ jet finder using charged tracks with minimum $\pT$ > 150 $\mc$.}
\includegraphics[width=0.55\textwidth]{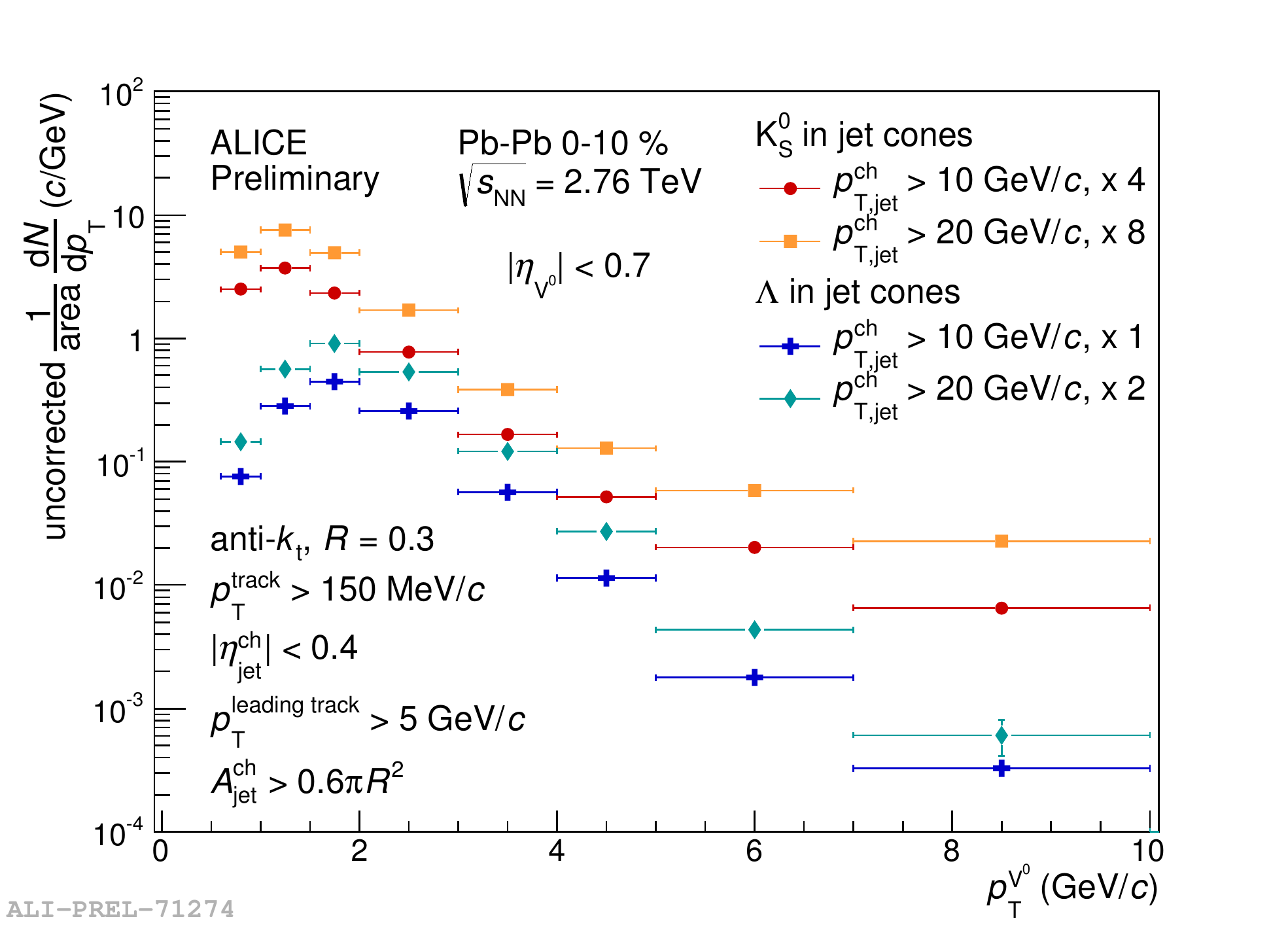}
\end{SCfigure}

Figure~\ref{fig:UncV0Jet} shows the uncorrected $\kos$ and $\lmb$ $\pT$ spectra for jets with R = 0.3 and for two different jet $\pT$ intervals ($\pTj$ > 10 $\gc$, > 20 $\gc$) in $\pbpb$ collisions. The spectra are scaled for better visibility. 

\begin{table}
\centering
\scriptsize

\begin{tabular}{|l|c|}
  \hline
  V0 UE estimation method & V0 sample  \\
  \hline
  No-jet events & in events without any selected jets (reconstructed jets fulfilling all further requirements)\\
  Outside cones & outside of 2R (R = 0.2,0.3)\\ 
  Random cones & in cones with randomly chosen axis ($\eta$,$\phi$) and no overlap with selected jets\\
  Median-cluster cones & uses median $\kt$ cluster (similar to $\kt$ alg. for average background estimation)\\
  Perpendicular cones & in cone perpendicular (in $\phi$) to jet axis\\
  \hline
\end{tabular}
\caption{Methods to estimate the V0 particle contribution from the UE}
\label{table:UEV0meth}
\end{table}

\begin{SCfigure}

\caption{\label{fig:UncV0Bulk} \addvbuffer[2.8ex 0ex] (top) Uncorrected $\kos$ $\pT$ spectra in UE in $\pbpb$ collisions at $\snn$ and in the 10$\%$ most central events. The different used methods serve to estimate systematic uncertainty of UE subtraction. (bottom) Each method divided by default (No-jet method), the ratios show a deviation of around 15$\%$  for $\pTv$ < 4 $\gc$.}
\centering
\includegraphics[width=0.55\textwidth]{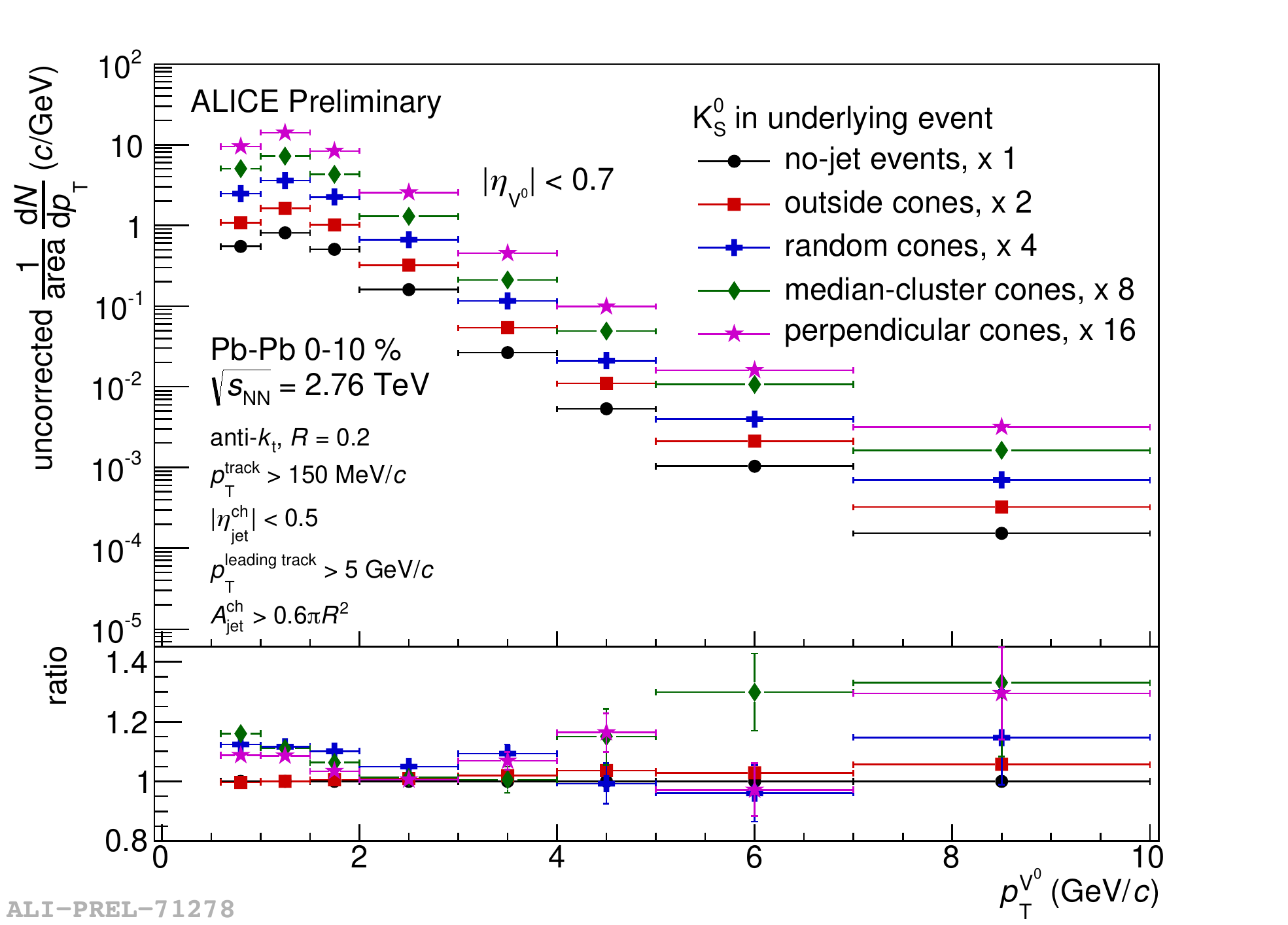}
\end{SCfigure}

Several corrections have to be applied to the raw yields. First, one has to correct for the V0 particle reconstruction efficiency as functions of $\pT$ and $\eta$, by using Monte-Carlo simulations. The efficiency is calculated separately for V0s in the inclusive event, the jet cone (JC) and in the UE. Since V0s from the UE contribute quite substantially to the V0 spectra in the JC, especially at low $\pT$, this has to be accurately subtracted. For this UE V0 correction, we developed several methods (see table \ref{table:UEV0meth}) to measure the raw V0 $\pT$ spectrum in the UE as shown in Figure~\ref{fig:UncV0Bulk}. The default method are inclusive V0s taken from events, in which no reconstructed jet is selected; this is the so-called "No-jet events" method. To assign a systematic uncertainty to this correction, 4 alternative methods are applied. Compared to the default method (see ratio in Figure~\ref{fig:UncV0Bulk}, bottom), they show a deviation of around 15$\%$  for $\pTv$ < 4 $\gc$. The order of the different correction steps is as follows: After the correction for the V0 reconstruction efficiency is applied, the UE V0 contribution (using the "no-jets" method) and the feed-down (FD) are subtracted from the V0 $\pT$ spectrum in the JC. To account for the sizable FD into $\lmb$ mostly stemming from $\Xi^{0,-}\decay\lmb$ and since there is no direct $\Xi^{0,-}$ measurement in jets so far, two extreme scenarios are considered for the V0s in the JC, namely the FD contribution from the inclusive particle analysis~\cite{LFspectrapaper} and the correction derived from PYTHIA pp simulations. The difference is used to assign a systematic uncertainty. 

\section{Results}
\begin{figure}[h]
\centering
\includegraphics[width=0.75\textwidth]{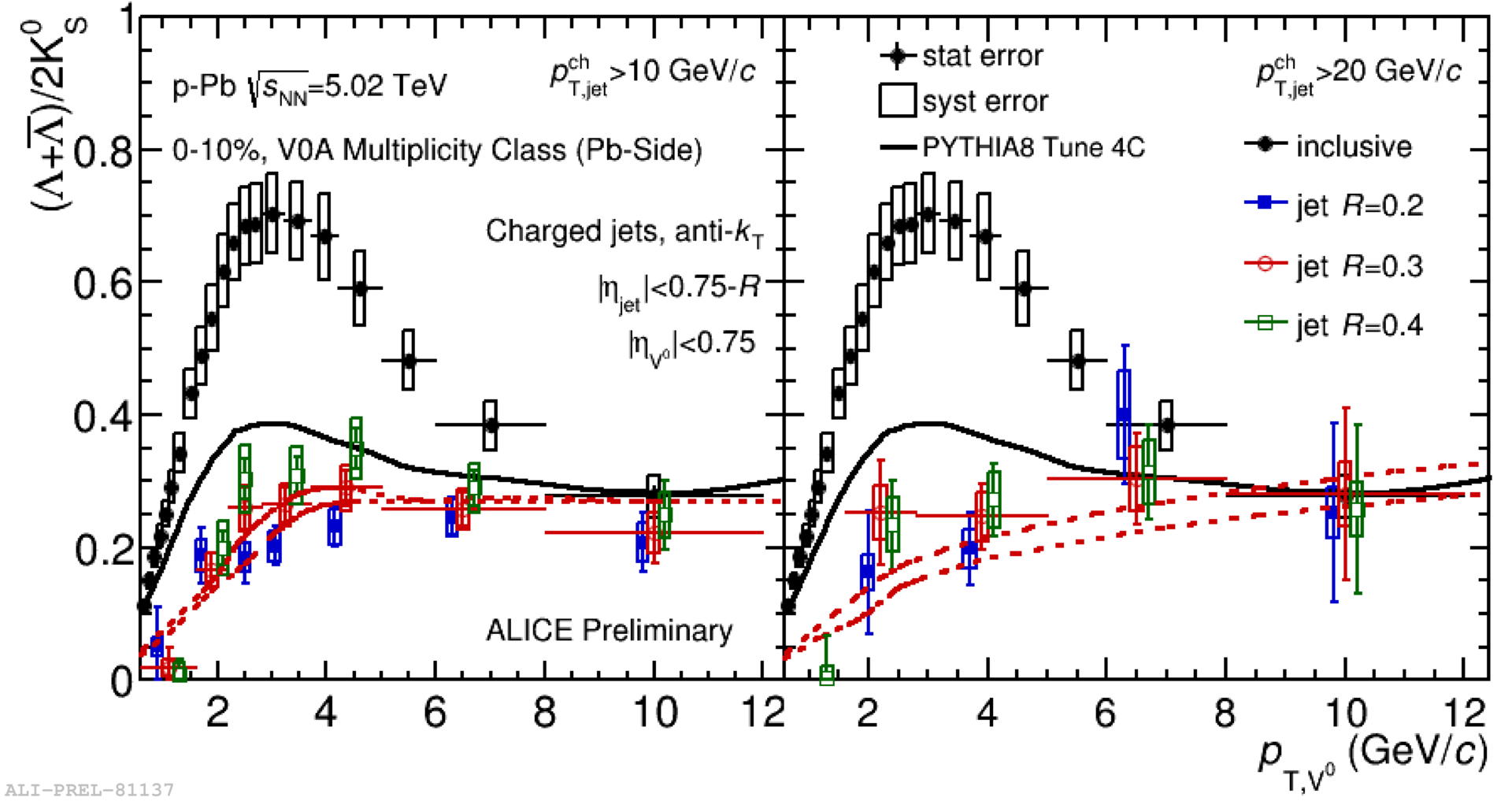}\\
\caption{\label{fig:RatiopPb} $\lmb$/$\kos$ ratio in jets for $\pTj$ > 10 $\gc$ (left) and  $\pTj$ > 20 $\gc$ (right) measured in $\ppb$ collisions at $\snnpPb$ for events in the 0-10$\%$ V0A (determined by the VZERO-A detector) multiplicity class~\cite{aliceexperiment, Abelev:2014ffa}.  Results of the ratio in jets are compared to the inclusive particle ratio (solid lines) and the expectation from a PYTHIA 8 4C simulation at 5.02 $\tc$ (dotted lines, showing spread for the used values of R).}
\end{figure}

Figure~\ref{fig:RatiopPb} shows the $\lmb$/$\kos$ ratio in jets in $\ppb$ events at 5.02 TeV and in the 10$\%$ highest multiplicity class. It is compared to the inclusive $\lmb$/$\kos$ ratio (solid lines) in $\ppb$ collisions and to the ratio predicted by PYTHIA 8 4C simulations  at 5.02 $\tc$ (dotted lines, showing spread for the used values of R). The measured ratio in jets shows no visible enhancement at intermediate $\pT$ and is far below the inclusive ratio. Within the systematic uncertainties, the ratio in jets is in agreement with the PYTHIA simulations. Furthermore there is no significant dependence on the $\pTj$ interval or on R visible. 

\section{Conclusions}
The observed $\lmb$/$\kos$ ratio in jets in $\ppb$ collisions is significantly smaller than the inclusive measurement and compatible with the predictions of a PYTHIA jet simulation. This points to a scenario in which the hadronisation of soft particles is disentangled from the hard fragmentation processes. Therefore the baryon enhancement seen in inclusive production (see black markers in Figure~\ref{fig:RatiopPb}) at intermediate particle $\pT$ in $\ppb$ collisions seems to be a feature of the UE and not from fragmentation. 
Furthermore, the uncorrected spectra for $\kos$ and $\lmb$ particles in the JC and in UE in 10$\%$ most central events in $\pbpb$ collisions have been presented. The $\lmb$/$\kos$ ratio in jets in $\pbpb$ collisions will be reported soon.

\section*{References}
\bibliographystyle{iopart-num}
\bibliography{bibliography.bib}

\end{document}